\documentstyle[amssymb,preprint,aps]{revtex}
\tightenlines

\begin{document}
\title{Crossover between polariton and phonon local states. Anisotropy-induced
localization threshold.}
\author{Victor Podolsky}
\address{Department of Physics, Queens College, CUNY, Flushing, NY 11367}
\date{\today}
\maketitle

\begin{abstract}
We consider the impurity-induced polariton local states in a dipole-active
medium. These states present the local optical vibrations which are
coherently coupled with the induced electromagnetic field. We show that the
crossover between the polariton and phonon local states takes place within
the relativistically narrow interval near the bottom of the polariton gap.
However, the resonant phonon-photon interaction leads to the absence of the
lower threshold of the impurity strength. These results are due to the
singularity of the density of the polariton states, which is generic for any
isotropic dipole-active phonon mode with the negative dispersion. We show
that a weak anisotropy removes this singularity and sets a finite threshold
for the local polariton states.
\end{abstract}


\section{Introduction}

The concept of the polariton local states was for the first time introduced
in Refs.\cite{Rupasov,John1}. Considering a two-level impurity atom
interacting with electromagnetic field in an isotropic frequency-dispersive
medium, the authors discovered the photon-atom bound states. These states
describe an excited atom which is coherently coupled with the medium
polaritons {\it via} the dipole interaction with the field. If the
intra-atom transition frequency is located inside the polariton gap, the
radiative relaxation of the bound states is suppressed, and the induced
electromagnetic field is localized around the impurity. It was shown that
the two-level model can be exactly solved by means of the Bethe-ansatz
technic. Authors constructed a complete set of the scattering and bound
states and analyzed a formation of the polariton-impurity band in a system
of spatially correlated impurities.

The results obtained in the above mentioned papers give a basis for the
theory of the photon localization in the crystals doped with the
dipole-active impurities. However, some questions need to be clarified. The
optical activity of the considered impurity atom was associated with its
electronic structure. In the case of the ``large atomic orbitals'', when the
characteristic scales of the polariton-atom bound states (coherence length,
localization radius) are comparable with the orbital size, the non-local
electronic structure of a defect should affect the results. In the opposite
case of the ``small orbitals'', the effects of the microscopic anisotropy of
the medium need to be taken into account.

Moreover, as it was shown in Ref.\cite{Leva}, {\it the spatial dispersion}
\thinspace of the medium has a surprisingly drastic effect on the polariton
local states. Considering a dipole-active medium with a point-like atomic
defect (an ``electronically rigid'' impurity ion, an isotope impurity,
etc.), authors found a new type of the defect-induced polariton local
states. These states are composed from phonons and photons localized around
the defect, and they are related to the phonon local states \cite
{Lifshitz,Maradudin}. However, unlike for pure phonon states, there is no
lower critical value of the impurity strength that needs to be exceeded for
the polariton local state to arise. It was shown that near the bottom of the
polariton gap the localization radius is macroscopically large, what
supports the long wavelength approximation used in \cite{Leva}.

In Ref.\cite{Cubic}, we analyzed the polariton local states associated with
non-isotope impurities. Considering a cubic crystal with a defect which
locally affects the crystal elastic constants, we found several local states
of different parity. All new states arise at the bottom of the polariton gap
upon infinitesimally small variations of the impurity parameters. This
feature of the polariton local states was attributed to the singularity of
the density of states at the bottom of the gap. It was shown that this
singularity is caused by the long-wavelength polaritons, what provides
applicability of the long-wavelength approximation to the local states near
the gap bottom.

In the present paper we investigate the crossover between the polariton and
phonon local states. The long-wavelength nature of the polariton states
allows us to analyze the crossover within the continuum model of a crystal.
In this approximation all details of ion displacements within elementary
cells are ignored and only resultant dipole momenta of the cells are
considered. It leaves the polarization and the electromagnetic fields as the
only relevant dynamical variables. Such approximation is appropriate when
the localization radius of the considered states greatly exceeds the lattice
parameter.

Our analysis shows that, in agreement with Ref.\cite{Leva}, a single
impurity embedded in the medium give rise to the local states in the
polariton gap. However, the frequency range available for the local states
is narrowed by the longitudinal band. We show that the states near the
bottom of the polariton gap are composed from the long-wavelength
polaritons. The typical momentum of these polaritons, $k_{\max },$ defines
the coherence length of the local states, $l_{%
\mathop{\rm coh}%
}\sim a\,\beta ^{-1/2}.$ The separation of the local states from the gap
bottom defines their localization radius, $l_{%
\mathop{\rm loc}%
}\sim a\beta ^{-1},$ where $a$ is the lattice parameter, and $\beta =v/c$ is
the ratio between the phonon's and photon's speeds. Despite the atomic size
of a defect, both characteristic lengths are macroscopic.

The long-wavelength nature of the polariton states allows us to analyze the
crossover between the polariton and phonon local states within the continuum
approximation. We show that the crossover takes place in the
relativistically narrow interval, $\Delta _{1}\!\sim \beta \Omega _{0}^{2},$
near the bottom of the polariton gap. \thinspace However, despite the small
size of the crossover region, the polariton resonance affects the selection
of the defects allowing local states in the entire gap and leads to the
absence of the\thinspace \thinspace lower threshold for the impurity
strength. This result is caused by the singularity of the density of the
polariton states, which is generic for any dipole-active phonon mode with
negative dispersion and isotropic spectrum. We showed that a weak crystal
anisotropy removes the singularity from the bottom and sets a finite
threshold for the local polariton states.

\section{Polariton local states}

Let us consider a dipole-active medium with an embedded impurity. Dynamical
equations of the medium can be introduced phenomenologically or derived from
the microscopic lattice equations. In the latter case the displacements of
ions within each elementary cell must be first expressed {\it via}\ the
dipole momentum of a cell, the displacement of its center of mass and other
similar variables, then the continuum limit of the equations must be worked
out. The resultant theory describes the long-wavelength dynamics of a
crystal where the electric field effectively interacts with\thinspace
\thinspace the optical phonons only: 
\begin{equation}
\left( \omega ^{2}-{\bf \hat{\Omega}}^{2}\right) {\bf {\cal P}}\!_{{\bf k}}=-%
\frac{d^{2}}{4\pi }\,{\bf E\,}_{{\bf k}}+\alpha \,\frac{a^{3}}{V}\,{\bf 
{\cal P}(0)},  \label{1}
\end{equation}
\begin{equation}
\left( \omega ^{2}-c^{2}k^{2}\right) \,{\bf E\,}_{{\bf k}}=-4\pi \left[
\left( \omega ^{2}-c^{2}k^{2}\right) \,{\bf \hat{P}}_{\Vert }+\omega ^{2}%
{\bf \hat{P}}_{\bot }\right] \,{\bf {\cal P}}_{\,{\bf k}}\,.  \label{2}
\end{equation}
Here ${\bf {\cal P}}\!_{{\bf k}}$ and ${\bf E\,}_{{\bf k}}$ are Fourier
components of the polarization and electric fields, ${\bf {\cal P}(0)}$ is
the defect polarization, ${\bf \hat{\Omega}}^{2}\left( {\bf k}\right) \ $is
the dynamical matrix of the medium, ${\bf \hat{P}}_{\Vert }$ and ${\bf \hat{P%
}}_{\bot }$ are the longitudinal and the transverse projectors in the
momentum space, $a^{\ }$is the lattice parameter, $V$ is the sample volume,
and $\ d^{2}$ is the photon-phonon coupling parameter.

Eq.$(1)$ implies that the defect is electrically identical with the replaced
host ion. In this case the ``defect strength'' can be expressed as\ $\alpha
=\omega ^{^{\prime \,}2}\delta \gamma /\gamma -\omega ^{2}\delta \mu /\mu ,$
where $\mu $ and $\gamma $ are the reduced mass of the elementary cell and
the elastic constant of the nearest-neighbor bonds, $\delta \mu $ and $%
\delta \gamma $ are their impurity-induced variations, and $\omega
^{^{\prime }\,2}$ is the characteristic frequency depending on the
properties of a defect in the crystal.

In the isotropic medium the dynamical matrix can be presented as follows: 
\begin{equation}
{\bf \hat{\Omega}}^{2}=\Omega _{\Vert }^{2}{\bf \hat{P}}_{\Vert }+\Omega
_{\bot }^{2}{\bf \hat{P}}_{\bot },  \label{3}
\end{equation}
where $\Omega _{\Vert }^{2}$ and $\Omega _{\bot }^{2}$ are the frequencies
of the longitudinal and transverse phonons, respectively. Below we assume a
negative dispersion in the phonon branches; so they have the following
long-wavelength asymptote: 
\begin{eqnarray}
\Omega _{\bot }^{2}\left( k\right) &\approx &\Omega _{0}^{2}-v_{\bot
}^{2}\,k^{2},  \label{4} \\
\Omega _{\Vert }^{2}\left( k\right) &\approx &\Omega _{\,0}^{2}-v_{\Vert
}^{2}\,k^{2},  \eqnum{$4^{\prime }$}
\end{eqnarray}
where the parameters $v_{\bot }$ and $v_{\Vert }$ set the ranges of the
typical phonon velocities, and $\Omega _{0}$ is the phonon activation
frequency. For the ``order of the magnitude'' estimates we assume $v_{\bot
}\sim v_{\Vert }\sim 10^{2}$m/s, \thinspace $\Omega _{0}\sim v/a\sim d,$ and 
$\alpha \sim \overline{\alpha }\ \Omega _{0}^{2},$ where $\overline{\alpha }$
is a numerical parameter.

Solving Eqs.$(1$-$3)$, one can express ${\bf E\,}_{{\bf k}}$ and ${\bf {\cal %
P}}\!_{{\bf k}}$ in terms of the defect polarization, ${\bf {\cal P}(0),}$
and obtain the spectral equation: 
\begin{equation}
1=\frac{\alpha }{3}\left( \frac{a}{2\pi }\right) ^{3}\int d{\bf k}\left[
\left( \omega ^{2}-\Omega _{\Vert }^{2}-d^{2}\right) ^{-1}+2\left( \omega
^{2}-\Omega _{\bot }^{2}-\frac{d^{2}\omega ^{2}}{\omega ^{2}-c^{2}k^{2}}%
\right) ^{-1}\right] =\alpha \,{I}(\omega ^{2})\,,  \label{5}
\end{equation}
where the integration is extended over the first Brillouin zone. This
equation defines spectra of all, extended and local, excitations. The
extended states form continuous bands and their dispersion relations are
determined by the poles of the integrand in Eq.$(5)$. In the isotropic
medium there is a single longitudinal branch, $\omega ^{2}=\Omega _{\Vert
}^{2}+d^{2},$ whereas the transverse band contains two polariton branches
[Fig.1]: 
\begin{equation}
\Omega _{\pm }\left( k\right) =\frac{1}{2}\left[ \sqrt{\left( \Omega _{\bot
}+ck\right) ^{2}+d^{2}}\pm \sqrt{\left( \Omega _{\bot }-ck\right) ^{2}+d^{2}}%
\right] .  \label{6}
\end{equation}
The lower branch, $\Omega _{-}\left( k\right) ,$ is activationless and
non-monotonic. Analysis of its ``large-momenta'' asymptote, 
\begin{equation}
\Omega _{-}\left( k\right) \approx \Omega _{\bot }\left( k\right) \left( 1-%
\frac{d^{2}}{2\,c^{2}k^{2}}\right) ,  \label{7}
\end{equation}
shows that the lower branch reaches its maximum at the point 
\begin{equation}
k_{\max }\approx \left( \frac{2\,d\Omega _{0}}{v_{\bot }c}\right) ^{1/2}\sim
\beta ^{1/2}a^{-1}.  \label{8}
\end{equation}

\thinspace Since $vk_{\max }\sim \beta ^{1/2}\Omega _{0}\ll \Omega _{0}\,\,$%
and $ck_{\max }\sim \beta ^{-1/2}\Omega _{0}\gg \Omega _{0},\,\,$the maximum
of $\Omega _{-}^{2}(k)\,\,\,$is located in the long-wavelength region but
far away from the cross-resonance point. Using Eqs.$(7$-$8)$ one can find: 
\begin{equation}
\Omega _{\max }^{2}=\Omega _{0}^{2}-\Delta _{1}\approx \Omega
_{0}^{2}-2\beta d\Omega _{0}.  \label{9}
\end{equation}

The polariton branches are separated by the gap which extends from $\Omega
_{\max }^{2}$ to $\!\Omega _{0}^{2}+d^{2}.$ \thinspace However, because the
longitudinal branch overlaps the top part of the polariton gap, the true
spectral gap, with no modes of any kind inside, is between $\Omega _{\max
}^{2}\,\,$and the minimum of the longitudinal branch, $\Omega _{\min
}^{2}=\Omega _{0}^{2}+d^{2}-\Delta _{\Vert }.$ \thinspace The spectral gap
is open only if the width of the polariton gap,\ $d^{2}+\Delta _{1},$
exceeds the width of the longitudinal band $\Delta _{\Vert }.$ Further below
we assume this and consider only local states inside the gap.

The frequency of the local state can be found from Eq.$(5)$. The
function\thinspace $\,I(\omega ^{2})$ in this equation presents a sum of the
``longitudinal'' and ``transverse'' terms, $I_{\Vert }(\omega ^{2})$ and $%
I_{\bot }(\omega ^{2}).$ The transverse integral can be written in the form: 
\begin{equation}
{I}_{\bot }(\omega ^{2})=\frac{2}{3}\left( \frac{a}{2\pi }\right) ^{3}\int 
\frac{\left( \omega ^{2}-c^{2}k^{2}\right) \,d{\bf k}}{\left( \omega
^{2}-\Omega _{+}^{2}\right) \left( \omega ^{2}-\Omega _{-}^{2}\right) }\!.
\label{10}
\end{equation}
When the frequency approaches $\Omega _{\max },$ this integral diverges at
the surface ${\bf k}^{2}=k_{\max }^{\,2}.$ Near the bottom of the gap $%
\left( \omega \gtrsim \Omega _{\max }\right) ,$ the transverse term
dominates in $I(\omega ^{2})$ and the region of $k\sim k_{\max }$ \thinspace
gives the major contribution to ${I}_{\bot }(\omega ^{2}).$ As it follows
from Eqs.$(4,9)$, in this region we can approximate: 
\begin{equation}
\Omega _{-}^{2}\left( k\right) \approx \Omega _{\max }^{2}-4\,\!v_{\bot
}^{2}\,\left( k-k_{\max }\right) ^{2}.  \label{11}
\end{equation}
It allows us to evaluate ${I}(\omega ^{2})$ in Eq.$(5)$ and obtain the
frequency of the local state: 
\begin{equation}
\sqrt{\omega ^{2}-\Omega _{\max }^{2}}\approx \frac{\alpha a}{3\pi v_{\bot }}%
\,\left( ak_{\max }\right) ^{2}\sim \overline{\alpha }\beta \Omega _{0}.
\label{12}
\end{equation}

In order to find the localization radius of the state we need to consider
the spatial distribution of the electric and the polarization fields. Using
the solution of Eqs.$(1$-$2)$, we obtain: 
\begin{equation}
{\bf E}\left( {\bf r}\right) =-\frac{\alpha a^{3}}{2\pi ^{2}}\int d{\bf k}%
\exp \left( i{\bf kr}\right) \left[ \frac{{\bf \hat{P}}_{\Vert }}{\omega
^{2}-\Omega _{\Vert }^{2}-d^{2}}+\frac{\omega ^{2}{\bf \hat{P}}_{\bot }}{%
\left( \omega ^{2}-\Omega _{+}^{2}\right) \left( \omega ^{2}-\Omega
_{-}^{2}\right) }\right] {\bf {\cal P}(0)}.  \label{13}
\end{equation}
Near the bottom of the gap the longitudinal field is small comparing to the
transverse term. Taking into account that the region of $k\sim k_{\max }$
\thinspace gives the main contribution to ${\bf E}_{\bot }\left( {\bf r}%
\right) $ and employing Eqs.$(11$-$12)$, one can find: 
\begin{equation}
{\bf E}_{\bot }\left( {\bf r}\right) \approx 4\pi \,\frac{\Omega _{0}^{2}}{%
c^{2}k_{\max }^{\,2}}\left\{ {\bf n\times }\left[ {\bf n\times {\cal P}(0)}%
\right] \right\} \,\frac{\sin \left( rk_{\max }\right) \exp \left(
-r\varkappa \right) }{rk_{\max }}.  \label{14}
\end{equation}
In a similar way one can obtain: 
\begin{equation}
{\bf {\cal P}}_{\bot }{\bf (r)}\approx \left\{ {\bf n\times }\left[ {\bf %
n\times {\cal P}(0)}\right] \right\} \,\frac{\sin \left( rk_{\max }\right)
\exp \left( -r\varkappa \right) }{rk_{\max }},  \label{15}
\end{equation}
where ${\bf n}$ is the unit radial vector, and 
\begin{equation}
\varkappa =\sqrt{\frac{\omega ^{2}-\Omega _{\max }^{2}}{4v_{\bot }^{2}}}%
\approx \frac{\alpha a}{6\pi v_{\bot }^{2}}\,\left( \,ak_{\max }\right)
^{2}\sim \overline{\alpha }\beta a^{-1}  \label{16}
\end{equation}
is the inverse localization radius of the considered state.

Equations $(14$-$15)$ allow us to estimate the energy partition between the
phonons and photons bound in the local state. Taking into account that the
polarization and the electric fields are concentrated within the $\varkappa
^{-1}$-range from the defect, we can evaluate their energies: 
\begin{eqnarray}
W_{mech} &\sim &\int {\bf {\cal P}}^{2}{\bf (r)}d{\bf r\sim {\cal P}}^{2}%
{\bf (0)}k_{\max }^{-2}\varkappa ^{-1},  \label{17} \\
W_{field} &\sim &\int {\bf E}^{2}{\bf (r)}d{\bf r\sim {\cal P}}^{2}{\bf (0)}%
\left( \frac{\Omega _{0}}{c\,k_{\max }}\right) ^{4}k_{\max }^{-2}\varkappa
^{-1}.  \eqnum{$17^{\prime }$}
\end{eqnarray}
It gives us the ``order of the magnitude'' estimate of the ratio of the
energies: 
\begin{equation}
\frac{W_{field}}{W_{mech}}\sim \left( \frac{\,\Omega _{0}}{ck_{\max }}%
\right) ^{4}\sim \beta ^{2}.  \label{18}
\end{equation}

\section{Crossover between the polariton and phonon local states}

The features of the polariton states near the gap bottom are caused by the
singularity of the density of states in the lower polariton band. Since $%
\Omega _{-}^{2}(k)$ reaches its maximum at the surface of a finite area
inside the Brillouin zone, the density of states diverges at $\Omega _{\max
} $: 
\begin{equation}
\rho \left( \omega ^{2}\right) =\left( \frac{a}{2\pi }\right) ^{3}\oint
d\sigma \left[ \frac{d\Omega _{-}^{2}(k)}{dk}\right] _{\Omega _{-}=\omega
}^{-1}\approx \frac{a(ak_{\max })^{2}}{\left( 2\pi \right) ^{2}v_{\bot }%
\sqrt{\Omega _{\max }^{2}-\omega ^{2}}}\,.  \label{19}
\end{equation}
This singularity is provided by the long-wavelength $\left( k\sim k_{\max
}\right) $ polaritons, which also dominate in the local states near the
bottom of the gap. It explains the macroscopic size of the coherence length, 
$l_{%
\mathop{\rm coh}%
}=k_{\max }^{-1}\sim a\beta ^{-1/2},$ and the localization radius, $l_{%
\mathop{\rm loc}%
}=\varkappa ^{-1}\sim a\beta ^{-1}.$ Also, because the typical momentum of
these modes is far away from the cross-resonance point $\left( k_{\max }\gg
k_{0}\right) $, the phonon content of the local states greatly exceeds their
photon content. The local polariton states appear first at the bottom of the
gap for $\alpha =+0$ and move inside the gap upon increase of the defect
strength. This further weakens the photon content of the polariton local
states and transforms them into the ordinary phonon states.

To investigate the crossover between the polariton and phonon local states
we need to analyze Eqs.$(1$-$2,5)$ in the entire spectral gap. Away from the
gap\ bottom both terms of Eq.$\left( 5\right) $, ${I}_{\Vert }(\omega ^{2})$
and $\,{I}_{\bot }(\omega ^{2}),$ become comparable$\ $and calculations here
require knowledge of the phonon dispersion laws $\Omega _{\Vert }^{2}(k)\ $%
and $\Omega _{\bot }^{2}(k)$ in the entire Brillouin zone. Such detailed
information is not consistent with the approximations employed in our model.
In the short-wavelength region a crystal anisotropy cannot be neglected, and
ion displacements within elementary cells must be considered in all details.
However, assuming that the crossover takes place near the gap bottom, we
proceed with calculations noting that in the isotropic model it is enough to
know the densities of states in the phonon bands. We use a simple
approximation accounting only for Kohn's singularities at the band
boundaries: 
\begin{equation}
\rho _{%
\mathop{\rm phon}%
}(\varepsilon )=\frac{8}{\pi \,\Delta _{%
\mathop{\rm phon}%
}^{2}}\sqrt{\left( \varepsilon -\omega _{\min }^{2}\right) \left( \omega
_{\max }^{2}-\varepsilon \right) },  \label{20}
\end{equation}
where $\Delta _{%
\mathop{\rm phon}%
}$ is the width of the phonon band; $\omega _{\min }^{2}\,$and $\omega
_{\max }^{2}$\thinspace \thinspace are the band boundaries.

To calculate ${I}_{\Vert }(\omega ^{2})$ we transform it into the integral
over the longitudinal band: 
\begin{equation}
{I}_{\Vert }(\omega ^{2})=\frac{1}{3}\left( \frac{a}{2\pi }\right) ^{3}\int 
\frac{d{\bf k}}{\omega ^{2}-\Omega _{\Vert }^{2}-d^{2}}=\frac{1}{3}%
\int_{\Omega _{\min }^{2}}^{\Omega _{\min }^{2}\!+\Delta _{\Vert }}\,\,\frac{%
\rho _{\Vert }(\varepsilon )\,d\varepsilon }{\omega ^{2}-\varepsilon },
\label{21}
\end{equation}
where $\Omega _{\min }^{2}=\Omega _{0}^{2}+d^{2}-\Delta _{\Vert }.\,$Using
Eq.$(20)$ for $\rho _{\Vert }(\varepsilon )$ one can obtain: 
\begin{equation}
{I}_{\Vert }(\omega ^{2})=-\,\frac{4}{3\Delta _{\Vert }^{2}}\left( \sqrt{%
\Omega _{\min }^{2}+\Delta _{\Vert }-\omega ^{2}}-\sqrt{\Omega _{\min
}^{2}-\omega ^{2}}\right) ^{2}.  \label{22}
\end{equation}

To evaluate the transverse integral we present it as a sum of two terms by
separating the Brillouin zone into two parts: 
\begin{equation}
{I}_{\bot }(\omega ^{2})=\frac{2}{3}\left( \frac{a}{2\pi }\right) ^{3}\int d%
{\bf k}\left( \omega ^{2}-\Omega _{\bot }^{2}-\frac{d^{2}\omega ^{2}}{\omega
^{2}-c^{2}k^{2}}\right) ^{-1}={I}_{\bot }^{\prime }(\omega ^{2})+{I}_{\bot
}^{\prime \prime }(\omega ^{2}).  \label{23}
\end{equation}
The first term, ${I}_{\bot }^{\prime }(\omega ^{2}),$ involves integration
over $k<k_{0}^{\prime }\,$ including the cross-resonance region, and ${I}%
_{\bot }^{\prime \prime }(\omega ^{2})$ presents the integral over the
remaining part of the Brillouin zone, $k>k_{0}^{\prime }.$ The separating
momentum $k_{0}^{\prime }$ is convenient to fix by the condition \ $\Omega
_{\bot }^{2}(k_{0}^{\prime })=\Omega _{\max }^{2}$ [Fig.1].

Since $k_{0}^{\prime }$ is located far away from the cross-resonance point,
the product $ck$ in ${I}_{\bot }^{\prime \prime }(\omega ^{2})$ greatly
exceeds the typical phonon frequencies and we can set there $ck=\infty $: 
\begin{equation}
{I}_{\bot }^{\prime \prime }(\omega ^{2})=\frac{2}{3}\left( \frac{a}{2\pi }%
\right) ^{3}\int_{k>k_{1}^{\prime }}\,\frac{d{\bf k}}{\omega ^{2}-\Omega
_{\bot }^{2}}=\frac{2}{3}\int_{\Omega _{0}^{2}-\Delta _{\bot }}^{\Omega
_{\max }^{2}}\,\frac{\rho _{\bot }(\varepsilon )\,d\varepsilon }{\omega
^{2}-\varepsilon }.  \label{24}
\end{equation}
Using Eq.$(20)$ for $\rho _{\bot }(\varepsilon ),$ we can calculate the last
integral: 
\begin{equation}
{I}_{\bot }^{\prime \prime }(\omega ^{2})=\frac{16}{3\pi \Delta _{\bot }}%
\left[ \left( 2\delta -1\right) \arcsin {\sqrt{\delta _{1}}}-\sqrt{\delta
_{1}\left( 1-\delta _{1}\right) }+\sqrt{\delta \left| 1-\delta \right| }%
\,F(\omega ^{2})\right] ,  \label{25}
\end{equation}
where 
\begin{equation}
F(\omega ^{2})=\left\{ 
\begin{array}{ll}
{+{\rm ln}\,\ {\frac{\sqrt{\delta \left( 1\!-\!\delta _{1}\right) }\!+\!%
\sqrt{\delta _{1}\left( 1\!-\!\delta \right) }}{\sqrt{\delta \left(
1\!-\!\delta _{1}\right) }\!-\!\sqrt{\delta _{1}\left( 1\!-\!\delta \right) }%
}}} & {{\rm for\ }\,\,\,\Omega _{\max }^{2}\leq \omega ^{2}\leq \Omega
_{0}^{2}\,,} \\ 
{-2\arctan {\sqrt{\frac{\delta _{1}\left( \delta \!-\!1\right) }{\delta
\left( 1\!-\!\delta _{1}\right) }}}} & {{\rm for}\,\,\,\,\omega ^{2}\!\geq
\Omega _{0}^{2}\,,}
\end{array}
\right.  \label{26}
\end{equation}
and we denote $\delta _{1}=(\Delta _{\bot }-\Delta _{1})\,/\Delta _{\bot },$
and $\ \delta =(\Delta _{\bot }+\omega ^{2}-\Omega _{0}^{2})\,/\,\Delta
_{\bot }.$

Considering $\omega ^{2}$ far away from the gap bottom $\left( \omega
^{2}-\Omega _{\max }^{2}\gg \Delta _{1}\right) ,$ one can obtain: 
\begin{equation}
{I}_{\bot }^{\prime \prime }(\omega ^{2})\approx \frac{8}{3\Delta _{\bot
}^{2}}\left( \sqrt{\omega ^{2}+\Delta _{\bot }-\Omega _{0}^{2}}-\sqrt{\omega
^{2}-\Omega _{0}^{2}}\right) ^{2}.  \label{27}
\end{equation}

In the opposite limit, $\omega \rightarrow \Omega _{\max },$ Eqs.$(25$-$26)$
lead to the logarithmic singularity: 
\begin{equation}
{I}_{\bot }^{\prime \prime }(\omega ^{2})\propto -\frac{16\Delta _{1}^{1/2}}{%
3\pi \Delta _{\bot }^{3/2}}{\rm ln}{\frac{\omega ^{2}-\Omega _{\max }^{2}}{%
\Delta _{\bot }}}.  \label{28}
\end{equation}

The second term in Eq.$(23)$ includes integration over the cross-resonance
region where the polariton effects cannot be neglected. Recalling that ${I}%
_{\bot }^{\prime }(\omega ^{2})$ presents the integral over the
long-wavelength modes and using Eqs.$\left( 4,20\right) $, we can rewrite ${I%
}_{\bot }^{\prime }(\omega ^{2})$ in the form: 
\begin{equation}
{I}_{\bot }^{\prime }(\omega ^{2})=\frac{2}{3}\int_{\Omega _{\max
}^{2}}^{\Omega _{0}^{2}}\frac{\left( \Omega _{0}^{2}-\varepsilon \right)
\,\rho _{\bot }(\varepsilon )\,\,d\varepsilon }{\left( \varepsilon -\omega
^{2}\right) \,\left( \varepsilon -\Omega _{0}^{2}\right) +\left( \omega
\,d\,v/c\right) ^{2}}=\frac{16\Delta _{1}}{3\pi \Delta _{\bot }^{2}}%
\int_{0}^{1}\frac{x^{3/2}\,\left( b-x\right) ^{1/2}\,dx}{\left(
x-a_{+}\right) \left( x-a_{-}\right) }.  \label{29}
\end{equation}
Here $b=\Delta _{\bot }\,/\,\Delta _{1}$ is a large parameter and two poles
of the integrand are given by the equation: 
\begin{equation}
a_{\pm }=\frac{1}{2\Delta _{1}}\left[ \ \Omega _{0}^{2}-\omega ^{2}\pm
\left( \omega ^{2}-\omega _{-}^{2}\right) ^{1/2}\left( \omega ^{2}-\omega
_{+}^{2}\right) ^{1/2}\right] ,  \label{30}
\end{equation}
where $\omega _{\pm }^{2}=\Omega _{0}^{2}\pm \Delta _{1},$ so that $\omega
_{-}^{2}$ coincides with the gap's bottom, $\Omega _{\max }^{2}.$

Direct calculations show that in the interval $\omega _{-}^{2}\leq \omega
^{2}\leq \omega _{+}^{2},$ where $a_{+}$ and $a_{-}$\ are complex-valued, ${I%
}_{\bot }^{\prime }(\omega ^{2})\ $has the form: 
\begin{eqnarray}
{I}_{\bot }^{\prime }(\omega ^{2}) &=&\frac{16\Delta _{1}^{1/2}}{3\pi \Delta
_{\bot }^{3/2}}\left[ 2+\frac{\nu ^{2}-3\eta ^{2}}{2\eta }\!\left( \arctan {%
\frac{1-\nu }{\eta }}+\arctan {\frac{1+\nu }{\eta }}\right) +\right. 
\nonumber \\
&&  \label{31} \\
&&\left. \frac{3\nu ^{2}-\eta ^{2}}{4\nu }\,{\rm ln}\,{\frac{(1-\nu
)^{2}+\eta ^{2}}{(1+\nu )^{2}+\eta ^{2}}}\right] \ ,  \nonumber
\end{eqnarray}
where the parameters $\nu $ and $\eta $ are defined by the equations: 
\begin{eqnarray}
&&\left. \nu ^{2}-\eta ^{2}=\frac{\Omega _{0}^{2}-\omega ^{2}}{2\Delta _{1}}%
,\right.  \label{32} \\
&&\left. 2\nu \eta =\frac{\left( \omega _{+}^{2}-\omega ^{2}\right)
^{1/2}\left( \omega ^{2}-\omega _{-}^{2}\right) ^{1/2}}{2\Delta _{1}}.\right.
\eqnum{$32^{\prime }$}
\end{eqnarray}
In agreement with the results obtained in the previous section, Eqs.$(31$-$%
32^{\prime })$ give the ``square-root singularity'' of ${I}_{\bot }^{\prime
}(\omega ^{2})$ at the gap bottom: 
\begin{equation}
{I}_{\bot }^{\prime }(\omega ^{2})\propto \frac{8\Delta _{1}}{3\Delta _{\bot
}^{3/2}}\left( \omega ^{2}-\Omega _{\max }^{2}\right) ^{-1/2}.  \label{33}
\end{equation}
For $\omega ^{2}\geq \Omega _{0}^{2}+\Delta _{1},$ when $a_{\pm }$ have real
negative values, we obtain: 
\begin{equation}
{I}_{\bot }^{\prime }(\omega ^{2})=\frac{32\Delta _{1}^{1/2}}{3\pi \Delta
_{\bot }^{3/2}}\left( 1-\frac{|a_{+}|^{3/2}\arctan {|}a{_{+}|^{-1/2}}%
-|a_{-}|^{3/2}\arctan {|}a{_{-}|^{-1/2}}}{|a_{+}|-|a_{-}|}\right) .
\label{34}
\end{equation}

Eqs.$(22,25$-$26,30$-$34)$ allow us to analyze the local states in the
entire gap. Since any local state presents a superposition of different
modes, we can qualitatively describe its composition comparing contributions
to ${I}(\omega ^{2})$ from different bands.

As we showed, the function ${I}(\omega ^{2})\ $has the singularity at the
gap bottom caused by the long-wavelength $\left( k\sim k_{\max }\right) $
polaritons. They also give the major contribution to ${I}(\omega ^{2})$ in
the interval, $\Omega _{\max }^{2}<\omega ^{2}\ll \Omega _{0}^{2}.$ However,
even at $\omega ^{2}=\Omega _{0}^{2}$ their contribution to ${I}(\omega
^{2}) $ is substantially weakened: 
\begin{equation}
{I}_{\bot }(\Omega _{0}^{2})=\frac{8}{3\Delta _{\bot }}\left[ 1+{\cal O}%
\left( \sqrt{\frac{\Delta _{1}}{\Delta _{\bot }}}\right) \right] \!,
\label{35}
\end{equation}
where the leading term is provided by the short-wavelength $\left(
k>k_{0}^{\prime }\right) $ polaritons, which are physically
indistinguishable from the transverse-optical phonons.

Outside the interval $\!\Omega _{\max }^{2}\leq \omega ^{2}\leq \Omega
_{\max }^{2}+2\Delta _{1}$ \ the function ${I}(\omega ^{2})\,\,$has
basically the phonon structure. All terms related to the polariton
singularity are weakened there by the power factors of the small parameter $%
\Delta _{1}/\Delta _{\bot }.$ At the upper boundary of the gap we obtain: 
\begin{equation}
{I}(\Omega _{\min }^{2})=\frac{8}{3\Delta _{\Vert }^{2}}\left[ \left( \frac{%
\sqrt{x+y\!}-\sqrt{y\!}}{x}\right) ^{2}-\frac{1}{2}\right] +{\cal O}\left( 
\sqrt{\frac{\Delta _{1}}{\Delta _{\bot }}}\right) ,  \label{36}
\end{equation}
where $x=\Delta _{\bot }\,/\,\Delta _{\Vert }$ and $y=d^{2}/\,\Delta _{\Vert
}-1.$

${I}(\Omega _{\min }^{2})$ determines the upper limit of the impurity
parameter which allows the local state to exist. Its value can be positive
or negative depending on the relations between the ion plasma frequency $%
d^{2}$ and the widths of phonon bands $\Delta _{\bot }$ and $\Delta _{\Vert
}.$ Analysis shows that ${I}(\Omega _{\min }^{2})\ $is positive only if the
width of the spectral gap does not exceed a certain value, $d^{2}-\Delta
_{\Vert }\leq (\Delta _{\Vert }\,/\,8)\left( 2-\Delta _{\bot }/\Delta
_{\Vert }\right) ^{2}.$\ Therefore, only in the ``narrow'' gaps all local
states are associated with impurities of the same type. In the case of the
isotope defects, for instance, the local states arise only in the presence
of light impurities. However, if $\ d^{2}-\Delta _{\Vert }$ exceeds the
critical value, the local states near the top and the bottom of the gap are
associated with the different types of defects. For isotopes those are heavy
and light impurities, respectively. The frequency regions of the
corresponding states are limited inside the gap. The upper frequency for the
``light'' states and lower frequency for the ``heavy'' ones are defined by
the equations, $\omega ^{2}{I}(\omega ^{2})=1$ and ${I}(\omega ^{2})=0,$
respectively.

\section{Anisotropy-induced threshold for the local states}

The considered above impurity-induced local states appear right at the
bottom of the polariton gap upon infinitesimally small variations of the
impurity parameters. This is in contrast with 3-D phonon systems where a
lower threshold for the local states always exists. The general theorem
regarding the existence of the threshold for local states in bandgaps of the
periodic systems was given in Ref.\cite{Figotin}. However, the proof
suggests that the density of states is regular everywhere. In our model, the
absence of the threshold is caused by the singularity of the density of
states in the lower polariton band, $\rho \left( \omega ^{2}\right) \propto
\left( \Omega _{\max }^{2}-\omega ^{2}\right) ^{-1/2}.\ $This singularity is
provided by the long-wavelength polaritons and is generic for any
dipole-active phonon mode with negative dispersion and isotropic spectrum.

However, even a weak crystal anisotropy can remove the singularity of the
density of states from the gap bottom. A dipole-active isotropic medium
presents the ``zero-order'' approximation of a\thinspace \thinspace cubic
crystal. The anisotropic terms of the phonon dispersion law in the cubic
crystals appear beyond the quadratic approximation: 
\begin{equation}
\Omega _{\bot }^{2}({\bf k})=\Omega _{0}^{2}-v_{\bot }^{2}k^{2}+\chi \Omega
_{0}^{2}\left( ak\right) ^{4}F\left( {\bf \hat{k}}\right)  \label{37}
\end{equation}
where $\chi $ is the small parameter, and $F\left( {\bf \hat{k}}\right) $ is
some anisotropic function. The anisotropic term, as one can see from Eq.$(7)$%
, makes the position $\left( \ k_{\max }\right) $ and the value of the
maxima in the lower polariton branch $\left( \ \Omega _{\max }^{2}\right) $
dependent on the crystallographic direction. In the case of a weak
anisotropy, the first effect can be neglected, what leads to the following
approximation of $\Omega _{-}^{2}\left( {\bf k}\right) $ near the surface $%
{\bf k}^{2}=k_{\max }^{2}:$%
\begin{equation}
\Omega _{-}^{2}\left( {\bf k}\right) \approx \Omega _{\max }^{2}\left[
1+\chi F\left( {\bf \hat{k}}\right) \left( ak_{\max }\right) ^{4}\right]
-4v^{2}\left( k-k_{\max }\right) ^{2}.  \label{38}
\end{equation}
The small magnitude of the anisotropic term, 
\begin{equation}
\left| \chi F\left( {\bf \hat{k}}\right) \left( ak_{\max }\right)
^{4}\right| \sim \chi \beta ^{2}\Omega _{\max }^{2}\ll \Omega _{\max }^{2},
\label{39}
\end{equation}
allows us to evaluate the asymptote of the density of states: 
\begin{equation}
\rho \left( \omega ^{2}\right) \varpropto \oint_{\Omega _{-}\left( {\bf k}%
\right) =\omega }d\sigma \left| \nabla _{{\bf k}}\Omega _{-}^{2}\left( {\bf k%
}\right) \right| ^{-1}\propto \left( \Omega _{\max }^{2}-\omega
^{2}+A\right) ^{-1/2},  \label{40}
\end{equation}
Since the crystal anisotropy destroys the continuous degeneration of the
polariton spectrum, it removes the singularity of the density of states from
the gap bottom. The anisotropy-induced shift of the singularity, 
\begin{equation}
A=\Omega _{\max }^{2}\chi \left( ak_{\max }\right) ^{4}\left\langle \left|
F\left( {\bf \hat{k}}\right) \right| \right\rangle ,  \label{41}
\end{equation}
where $\left\langle \left| F\left( {\bf \hat{k}}\right) \right|
\right\rangle \sim 1$ is the ``average'' value of the anisotropic function,
defines a finite threshold for the polariton local states: 
\begin{equation}
\alpha _{\min }\sim \Omega _{0}A^{1/2}\sim \chi ^{1/2}\beta ^{2}\Omega
_{0}^{2}.  \label{42}
\end{equation}

\section{Summary and conclusions}

We considered the local states in an isotropic dipole-active medium. It was
shown that, in agreement with Ref.\cite{Leva}, a single impurity embedded in
the medium give rise to a local state in the polariton gap. The frequency
interval available for the local states is extended from the top of the
lower polariton band, $\Omega _{\max }^{2},$ to the bottom of the
longitudinal band.

The polariton local states appear right at the bottom of the polariton gap
for an infinitesimally small value of the impurity strength. Our analysis
showed that the local states near $\,\Omega _{\max }^{2}$ are composed from
the long-wavelength polaritons. The typical momentum of these polaritons, $%
k_{\max },$ defines the coherence length of the local states, $l_{%
\mathop{\rm coh}%
}\sim a\,\beta ^{-1/2}$. The separation of the local states from the gap
bottom, $\sqrt{\omega ^{2}-\Omega _{\max }^{2}},$ defines their localization
radius, $l_{%
\mathop{\rm loc}%
}\sim a\,\beta ^{-1}\overline{\alpha }^{-1}.$ Despite the atomic size of a
defect, both characteristic lengths are macroscopic near the gap bottom.

These results, and the absence of the\thinspace \thinspace lower threshold
for the impurity strength, are due to the singularity of the density of the
polariton states, $\rho _{-}\left( \omega ^{2}\right) \propto \left( \Omega
_{\max }^{2}-\omega ^{2}\right) ^{-1/2}.\ $This singularity is caused by the
long-wavelength polaritons and is generic for any dipole-active phonon mode
with negative dispersion and isotropic spectrum. We showed that a weak
crystal anisotropy removes the singularity from the bottom, $\rho _{-}\left(
\omega ^{2}\right) \propto \left( \Omega _{\max }^{2}-\omega ^{2}+A\right)
^{-1/2},$ and sets a finite threshold for the local polariton states, $%
\alpha _{\min }\sim \Omega _{0}\,A^{1/2}.$

The long-wavelength nature of the polariton states allowed us to analyze the
crossover between the polariton and phonon local states within the continuum
approximation. We found that the crossover takes place in the
relativistically narrow interval, $\Delta _{1}\!\sim \beta \Omega _{0}^{2},$
near the bottom of the polariton gap. \thinspace However, despite the small
size of the crossover region, the polariton resonance affects the selection
of the defects allowing local states in the entire gap. We showed that in
the ``narrow'' gaps there must be only light isotope impurities.

In the presence of a macroscopic concentration of impurities, the local
states can provide the transmission inside the spectral gap. Since $l_{%
\mathop{\rm loc}%
}$ greatly exceeds $l_{%
\mathop{\rm coh}%
},$ the transmission regime critically depends on the impurity
concentration, $n.$ When $n^{-1/3}\ll $ $l_{%
\mathop{\rm loc}%
}$ \thinspace the probability of the tunneling of excitations from one
impurity to another is exponentially small, what corresponds to the
localized regime. For $l_{%
\mathop{\rm coh}%
}<n^{-1/3}\lesssim $ $l_{%
\mathop{\rm loc}%
}$ the resonant tunelling gives rise to the diffuse propagation of the
radiation. When $n^{-1/3}$ becomes less than $l_{%
\mathop{\rm coh}%
},$ the polariton-impurity band begins to form, and the transmission regime
regains properties of the coherent propagation.

The local polariton states provide a ``phonon assisted localization'' of
electromagnetic waves. However, our estimate of the energy partition shows a
strong suppression of the photon content of these states, $%
W_{field}/W_{mech}\sim \beta ^{2}$. It is caused by the fact that $k_{\max }$
is much greater than the cross-resonance momentum $k_{0}$. To eliminate this
disproportion, we need to lower the group velocity of electromagnetic waves
in the active medium. It can be achieved in the medium placed inside a
waveguide. For instance, the dispersion law of the modes propagating in the
parallel-plate waveguide, $\omega /c=\sqrt{(\pi n/l)^{2}+k^{2}},$ provides
the reduction of their group velocity in the long-wavelength region.
Estimates show that for $l\sim 10^{6}a$ the phonon and photon velocities
become comparable in the cross-resonance region. The detailed analysis of
the possibility to obtain the local modes in a thin waveguide filled with a
dipole-active medium will be presented elsewhere.

\section*{Acknowledgments}

We wish to thank L.I. Deych and A.A. Lisyansky for useful discussion and
comments on the manuscript. This work was partially supported by a CUNY
collaborative grant.

\end{document}